\documentclass[aps,prl,twocolumn,showpacs,superscriptaddress,groupedaddress,nofootinbib]{revtex4-2}  
\usepackage{cancel}
\usepackage{graphicx} 
\usepackage[T1]{fontenc}
\usepackage{lmodern} 
\usepackage{graphicx}  
\usepackage{dcolumn}   
\usepackage{bm}        
\usepackage{amssymb}   
\usepackage{amsmath}
\usepackage{bbold}
\usepackage{array}
\usepackage{float}
\usepackage{makecell}
\usepackage{braket}
\usepackage{longtable}
\usepackage{supertabular,booktabs}
\usepackage{soul}

\usepackage{titlesec} 
\usepackage[colorlinks,citecolor=blue,urlcolor=blue,hypertexnames=true]{hyperref}
\usepackage{subfigure}

\usepackage[usenames,dvipsnames,svgnames,table]{xcolor} 

\newcommand{\be}{\begin{equation}}
\newcommand{\ee}{\end{equation}}
\newcommand{\bea}{\begin{eqnarray}}
\newcommand{\eea}{\end{eqnarray}}
\newcommand{\bml}{\begin{subequations}}
\newcommand{\eml}{\end{subequations}}
\newcommand{\bfig}{\begin{figure}}
\newcommand{\efig}{\end{figure}}

\newcommand{\bmat}{\begin{pmatrix}}
\newcommand{\emat}{\end{pmatrix}}
\usepackage{graphicx, slashed,booktabs, color, multirow, float,
amsfonts, bbold, mathtools, sidecap, tikz, bm,enumitem}
\usepackage{multirow}
\usepackage{bbding}
\usepackage{amsmath}
\usepackage{titlesec}
\usepackage{hyperref}
\usepackage{wasysym}
\usepackage{amssymb}
\usepackage{pifont}

\usepackage[dvipsnames, usenames]{xcolor}

\definecolor{linkcolor}{rgb}{0.55, 0.13, .32}

\definecolor{oucrimsonred}{rgb}{0.6, 0.0, 0.0}
\definecolor{persianblue}{rgb}{0.11, 0.22, 0.73}
\definecolor{forestgreen}{rgb}{0.13,0.35,0.13}
\definecolor{lightgray}{rgb}{0.83, 0.83, 0.83}
 \hypersetup{colorlinks, citecolor=oucrimsonred, linkcolor=persianblue, urlcolor=oucrimsonred}
\definecolor{cornellred}{rgb}{0.7, 0.11, 0.11}
\definecolor{navyblue}{rgb}{0.0, 0.0, 0.5}
\definecolor{amethyst}{rgb}{0.6, 0.4, 0.8}
\definecolor{yellow}{rgb}{1.0, 1.0, 0.0}
\definecolor{firebrick}{rgb}{0.7, 0.13, 0.13}
\definecolor{tangerineyellow}{rgb}{1.0, 0.8, 0.0}
\definecolor{deepfuchsia}{rgb}{0.76, 0.33, 0.76}
\definecolor{amber}{rgb}{1.0, 0.75, 0.0}
\definecolor{VioletRed4}{rgb}{0.55, 0.13, .32}
\definecolor{indiagreen}{rgb}{0.07, 0.53, 0.03}
\definecolor{VioletRed4}{rgb}{0.55, 0.13, .32}

\usepackage{hyperref}
\usepackage{graphics, appendix, afterpage, makecell} 
\usepackage{bbold}
\usepackage{tikz}
\usepackage{adjustbox}

\usepackage{tcolorbox}

\definecolor{oucrimsonred}{rgb}{0.6, 0.0, 0.0}
\definecolor{persianblue}{rgb}{0.11, 0.22, 0.73}
\definecolor{forestgreen}{rgb}{0.13,0.35,0.13}
\definecolor{lightgray}{rgb}{0.83, 0.83, 0.83}
 \hypersetup{colorlinks, citecolor=oucrimsonred, linkcolor=persianblue, urlcolor=oucrimsonred}
\definecolor{cornellred}{rgb}{0.7, 0.11, 0.11}
\definecolor{navyblue}{rgb}{0.0, 0.0, 0.5}
\definecolor{amethyst}{rgb}{0.6, 0.4, 0.8}
\definecolor{yellow}{rgb}{1.0, 1.0, 0.0}
\definecolor{firebrick}{rgb}{0.7, 0.13, 0.13}
\definecolor{tangerineyellow}{rgb}{1.0, 0.8, 0.0}
\definecolor{deepfuchsia}{rgb}{0.76, 0.33, 0.76}
\definecolor{amber}{rgb}{1.0, 0.75, 0.0}
\definecolor{VioletRed4}{rgb}{0.55, 0.13, .32}
\definecolor{indiagreen}{rgb}{0.07, 0.53, 0.03}
\definecolor{VioletRed4}{rgb}{0.55, 0.13, .32}

\definecolor{oucrimsonred}{rgb}{0.6, 0.0, 0.0}
\newcommand\vertarrowbox[3][6ex]{%
  \begin{array}[t]{@{}c@{}} #2 \\
  \left\uparrow\vcenter{\hrule height #1}\right.\kern-\nulldelimiterspace\\
  \makebox[0pt]{\scriptsize#3}
  \end{array}%
}

\definecolor{mtcolor}{rgb}{.8,.3,.1}

\definecolor{violachiaro}{rgb}{1,0.6,1}

\definecolor{gbcolor}{rgb}{.43,.22,.12}
 
\definecolor{gbcolor2}{rgb}{.9,.2,.6}
\definecolor{gbcolor3}{rgb}{.3,.2,.6}

\definecolor{verdechiaro}{rgb}{0.6,1,0.6}
\definecolor{giallochiaro}{rgb}{1,1,0.6}
\definecolor{bluscuro}{rgb}{0.15, 0.2, 0.9}
\definecolor{verdes}{rgb}{0.1, 0.5, 0.1}%
\definecolor{tangerineyellow}{rgb}{1.0, 0.8, 0.0}
\definecolor{smokyblack}{rgb}{0.06, 0.05, 0.03}

\definecolor{americanrose}{rgb}{1.0, 0.01, 0.24}
\definecolor{cobalt}{rgb}{0.0, 0.28, 0.67}
\definecolor{brandeisblue}{rgb}{0.0, 0.44, 1.0}
\definecolor{mycolor}{rgb}{0.0, 0.0, 0.5}
\definecolor{oxfordblue}{rgb}{0.0, 0.13, 0.28}
\definecolor{azure}{rgb}{0.0, 0.5, 1.0}
\definecolor{turquoiseblue}{rgb}{0.0, 1.0, 0.94}
\newtcolorbox{mynewbox}[1]{colback=white!5!white,colframe=azure!75!black,fonttitle=\bfseries,title=#1}
\newtcolorbox{mybox}{colback=mycolor!5!white,colframe=azure!75!black}
\newtcolorbox{mynamedbox}[1]{colback=mycolor!5!white,colframe=azure!75!black,title=#1}
\definecolor{venetianred}{rgb}{0.78, 0.03, 0.08}
\newtcolorbox{mynamedbox1}[1]{colback=venetianred!5!white,colframe=venetianred!80!black,title=#1}
\newtcolorbox{mynamedbox2}[1]{colback=azure!5!white,colframe=azure!80!black,title=#1}

\definecolor{rossocorsa}{rgb}{0.83, 0.0, 0.0}

\tikzset{->-/.style={decoration={
  markings,
  mark=at position #1 with {\arrow{>}}},postaction={decorate}}}
\tikzset{-<-/.style={decoration={
  markings,
  mark=at position #1 with {\arrow{<}}},postaction={decorate}}} 

\def\be{\begin{equation}}
\def\ee{\end{equation}}
\def\ba{\begin{eqnarray}}
\def\ea{\end{eqnarray}}

\def\L*{{\cal L}_*}
\def\L{\mathcal{L}}
\def\({\left(}
\def\){\right)}

\def\<{\langle}
\def\>{\rangle}

\def\cs2{c_{s}^{2}}

 \def\be   {\begin{equation}}   \def\ee   {\end{equation}}
 \def\ba   {\begin{array}}      \def\ea   {\end{array}}
 \def\bea  {\begin{eqnarray}}   \def\eea  {\end{eqnarray}}
 \def\bean {\begin{eqnarray*}}  \def\eean {\end{eqnarray*}}

\titleclass{\subsubsubsection}{straight}[\subsection]

\newcounter{subsubsubsection}[subsubsection]
\renewcommand\thesubsubsubsection{\thesubsubsection.\arabic{subsubsubsection}}

\titleformat{\subsubsubsection}
  {\normalfont\normalsize\bfseries}{\thesubsubsubsection}{1em}{}
\titlespacing*{\subsubsubsection}
{0pt}{3.25ex plus 1ex minus .2ex}{1.5ex plus .2ex}

\makeatletter
\renewcommand\paragraph{\@startsection{paragraph}{5}{\z@}%
  {3.25ex \@plus1ex \@minus.2ex}%
  {-1em}%
  {\normalfont\normalsize\bfseries}}
\renewcommand\subparagraph{\@startsection{subparagraph}{6}{\parindent}%
  {3.25ex \@plus1ex \@minus .2ex}%
  {-1em}%
  {\normalfont\normalsize\bfseries}}
\def\toclevel@subsubsubsection{4}
\def\toclevel@paragraph{5}
\def\toclevel@paragraph{6}
\def\l@subsubsubsection{\@dottedtocline{4}{7em}{4em}}
\def\l@paragraph{\@dottedtocline{5}{10em}{5em}}
\def\l@subparagraph{\@dottedtocline{6}{14em}{6em}}
\makeatother

\usepackage{titlesec} 
\usepackage[colorlinks,citecolor=blue,urlcolor=blue,hypertexnames=true]{hyperref}
\usepackage{subfigure}

\usepackage[usenames,dvipsnames,svgnames,table]{xcolor}

\usepackage{tikzsymbols}
\usepackage{natbib}
\usepackage{float}

\usepackage{tikz,xcolor,hyperref}

\usepackage{slashed}

\definecolor{lime}{HTML}{A6CE39}
\DeclareRobustCommand{\orcidicon}{
	\begin{tikzpicture}
	\draw[lime, fill=lime] (0,0) 
	circle [radius=0.2] 
	node[white] {{\fontfamily{qag}\selectfont \tiny ID}};
	\draw[white, fill=white] (-0.0625,0.095) 
	circle [radius=0.007];
	\end{tikzpicture}
	\hspace{-2mm}
}

\usepackage[usenames,dvipsnames,svgnames,table]{xcolor}  
\usepackage{tikzsymbols}
\usepackage{natbib}
\usepackage{float}

\usepackage{tikz,xcolor,hyperref}

\usepackage{slashed}

\definecolor{lime}{HTML}{A6CE39}
\DeclareRobustCommand{\orcidicon}{
	\begin{tikzpicture}
	\draw[lime, fill=lime] (0,0) 
	circle [radius=0.2] 
	node[white] {{\fontfamily{qag}\selectfont \tiny ID}};
	\draw[white, fill=white] (-0.0625,0.095) 
	circle [radius=0.007];
	\end{tikzpicture}
	\hspace{-2mm}
}

\foreach \x in {A, ..., Z}{\expandafter\xdef\csname orcid\x\endcsname{\noexpand\href{https://orcid.org/\csname orcidauthor\x\endcsname}
			{\noexpand\orcidicon}}
}

\usepackage{bbold}
\usepackage{tikz}
\usepackage{adjustbox}
\usepackage{tcolorbox}
\usepackage{enumitem}
\usepackage{amsfonts}

\setlist[itemize,1]{label=$\times$}
\setlist[itemize,2]{label=$\checkmark$}
\setlist[itemize,3]{label=$\diamond$}
\setlist[itemize,4]{label=$\bullet$}

\begin{document}
\title{\Large \textcolor{Sepia}{Quintessential Inflation in Light of ACT DR6}}
\author{\large Sayantan Choudhury\orcidA{}${}^{1}$}
\email{sayantan_ccsp@sgtuniversity.org, \\
sayanphysicsisi@gmail.com, \\
sayantan.choudhury@nanograv.org (Corresponding author)} 

\author{\large Swapnil Kumar Singh\orcidF{}${}^{2}$}
\email{swapnil.me21@bmsce.ac.in, swapnilsingh.ph@gmail.com}

\author{\large Satish Kumar Sahoo\orcidG{}${}^{3}$}
\email{satishsahoophysics@gmail.com}


\affiliation{${}^{1}$ Centre for Cosmology and Science Popularization (CCSP), SGT University, Gurugram, Delhi- NCR, Haryana- 122505, India.}

\affiliation{ ${}^{2}$B.M.S. College of Engineering, 
    Bangalore, Karnataka, 560019, India.}

\affiliation{${}^{3}$National Institute of Technology Rourkela, Rourkela 769008, Odisha, India.}

\begin{abstract}
We perform a precision investigation of smooth quintessential inflation in which a single canonical scalar field unifies the two known phases of cosmic acceleration. Using a CMB-normalized runaway exponential potential, we obtain sharply predictive inflationary observables: a red-tilted spectrum with $n_s = 0.964241$ and an exceptionally suppressed tensor-to-scalar ratio $r = 7.48 \times 10^{-5}$ at $N=60$, lying near the optimal region of current Planck+ACT constraints. Remarkably, all observable scales exit the horizon within an extremely narrow field interval $\Delta\phi \simeq 0.03\,M_{\rm Pl}$, tightly linking early- and late-time dynamics and reducing theoretical ambiguities. While inflationary tensors remain invisible to CMB B-mode surveys, the subsequent stiff epoch—an intrinsic hallmark of quintessential cosmology—imprints a blue-tilted stochastic gravitational-wave background within the discovery reach of future interferometers such as LISA, DECIGO, ALIA, and BBO. Our results demonstrate that this minimal, featureless model not only survives current bounds, but provides concrete, falsifiable predictions across gravitational-wave frequencies spanning over twenty orders of magnitude.
\end{abstract}

\maketitle
\section{Introduction}

Inflation remains the most compelling and internally consistent paradigm for the physics of the primordial Universe. It provides a dynamical mechanism for addressing the shortcomings of the standard hot Big Bang scenario, including the horizon, flatness and monopole problems~\cite{Guth:1980zm,Linde:1981mu,Baumann:2009ds,Senatore:2016aui,Choudhury:2024aji,Kallosh:2025ijd,Odintsov_2023}, and exponentially suppresses any primordial spatial curvature, driving the cosmos toward an almost perfectly homogeneous and isotropic state on large scales. As a further consequence of accelerated expansion, quantum fluctuations of the inflaton and the metric become imprinted as primordial scalar and tensor perturbations~\cite{Mukhanov:1981xt}. These perturbations subsequently evolve into the observed Cosmic Microwave Background (CMB) anisotropies and into the distribution of matter that forms the large-scale structure (LSS) of the Universe.

An extensive body of CMB observations has transformed inflation from a broad phenomenological hypothesis into a quantifiable framework. The combined results from Planck, WMAP, ACT and SPT~\cite{Hinshaw2013WMAP,Planck:2018jri,ACT:2025fju,ACT:2025tim,Balkenhol2023SPT3G} have constrained the scalar spectral index to remarkable precision, enabling a highly nontrivial test of inflationary model space. Current analyses suggest the preference for a red-tilted power spectrum, while showing indications of an upward directional shift in the inferred spectral tilt. For instance, the latest ACT+Planck (P--ACT) datasets yield $n_s = 0.9709 \pm 0.0038$, while incorporating DESI DR2 clustering further tightens $n_s = 0.9743 \pm 0.0034$ together with $r < 0.038$ at $95\%$~CL~\cite{ACT:2025fju,ACT:2025tim, DESI:2024mwx,DESI:2024uvr}. This refinement is of particular theoretical significance\footnote{A shift of even a few parts in $10^{-3}$ can exclude entire universality classes of single–field slow–roll models due to their strong predictive correlations between $n_s$ and $r$.} because many models that comfortably matched Planck-only constraints now lie at the periphery of the updated confidence intervals. As a result, a large number of recent studies have focused on improving or extending inflationary dynamics via nonminimal gravitational couplings, higher-derivative sectors, or modified attractor behaviors~\cite{Ellis:2025ieh,Antoniadis:2025arXiv,Gialamas:2025arXiv,Addazi:2025arXiv,Mohammadi:2025arXiv,Haque:2025arXiva,Haque:2025arXivb,Drees:2025arXiv,Kallosh:2025arXiv,Aoki:2025arXiv,Mondal:2025arXiv,Liu:2025arXiv,Maity:2025arXiv,Gao:2025arXiv,Dioguardi:2025arXiv,McDonald:2025arXiv,Yin:2025arXiv,Yi:2025arXiv,Peng:2025arXiv,He:2025arXiv,Wolf:2025arXiv,Berera:2025arXiv,Pallis:2025nrv,Odintsov:2025bmp,Zhu:2025twm,Ketov:2025cqg,Zahoor:2025nuq,Chakraborty2025_BehaviourAlphaAttractors_WarmInflation,Han2025_HiggsPoleInflation_ACT,Gao2025_ObservationalConstraints_NMC_ACT,Choudhury2025_NewPhysics_ACT_DESI,Yi2025_PolynomialAlphaAttractor,He2025_IncreaseNS_PoleInflation_EC,Heidarian:2025drk,Oikonomou:2025xms,Yuennan:2025kde,Oikonomou:2025htz,Odintsov:2025jky,Singh:2025uyr}.

Although the CMB provides exquisite information at scales $k \lesssim \mathcal{O}(1~\mathrm{Mpc}^{-1})$, its reach does not extend to smaller scales where inflation may depart significantly from smooth slow-roll behavior. Consequently, a major frontier in inflationary cosmology has emerged: the exploration of small-scale primordial perturbations through gravitational wave observations and related probes. Recent pulsar timing array (PTA) measurements from NANOGrav, EPTA, PPTA and CPTA~\cite{NANOGrav:2023gor,NANOGrav:2023hde,EPTA:2023fyk,Reardon:2023zen,Xu:2023wog} strongly support the existence of a nanohertz stochastic gravitational-wave background (SGWB). While astrophysical sources remain a plausible interpretation, an origin in the early Universe\footnote{Second-order gravitational waves sourced by scalar perturbations are unavoidable in General Relativity once the linear scalar fluctuations become large on subhorizon scales.} has gained increasing traction owing to modest spectral deviations from the canonical supermassive black hole binary expectation. In particular, second-order induced gravitational waves (IGWs) produced during radiation domination can account for the observed PTA signals if the scalar power spectrum is sufficiently enhanced on scales around $k \sim 10^{6}$--$10^{8}~\mathrm{Mpc}^{-1}$~\cite{NANOGrav:2023hvm,Franciolini:2023pbf}. Localized features in the inflationary potential, including Gaussian bumps and near-inflection points, provide a natural mechanism for this enhancement~\cite{Atal:2019cdz,Mishra:2019pzq}, without violating CMB-scale slow-roll constraints. The same overdensities may additionally lead to the formation of primordial black holes (PBHs)~\cite{Carr:2016drx}, providing a unified connection between early-Universe inflationary dynamics and dark matter phenomenology.

Among early–Universe models that simultaneously address high-redshift inflation and low-redshift acceleration, quintessential inflation holds a unique theoretical position~\cite{Spokoiny:1993kt,Peebles:1998qn,Sahni:2001qp,Sami:2004ic,Hossain:2014zma}. Here, a single canonical scalar field with a runaway potential is responsible for both the initial accelerated expansion and the late-time dark-energy–dominated epoch. Following the end of inflation, the steep descent of the potential naturally triggers a kination regime in which the energy density scales as $\rho_\phi \propto a^{-6}$ and the equation of state approaches $w \simeq 1$. Conventional reheating via inflaton oscillations is therefore absent and particle production must occur through non-oscillatory channels such as instant preheating or gravitational particle creation~\cite{Bassett:2005xm,Hossain:2014xha}. The duration of the stiff epoch is constrained by Big Bang Nucleosynthesis (BBN), since gravitational waves entering the horizon during kination are amplified relative to those entering during radiation domination~\cite{Figueroa:2018twl,Kuroyanagi:2008ye}. The resulting primordial tensor background acquires a strongly blue-tilted spectral shape at high frequencies~\cite{Giovannini:1999qj,Riazuelo:2000fc,Ahmad:2019jbm}, placing it within the prospective reach of LISA, DECIGO, ALIA and BBO.

The interplay between transient violations of slow-roll caused by localized features and the post-inflationary stiff expansion phase leads to a distinctive multi-frequency observational signature: an enhanced and narrow-band IGW component in the PTA range accompanied by a blue-tilted ultraviolet tail at interferometer scales. This embedding of early-Universe gravitational-wave phenomenology offers a means of simultaneously testing the inflationary potential, reheating history, and the nature of late-time acceleration within a single field-theoretic framework.

\begin{figure*}
\centering
\includegraphics[width=\linewidth]{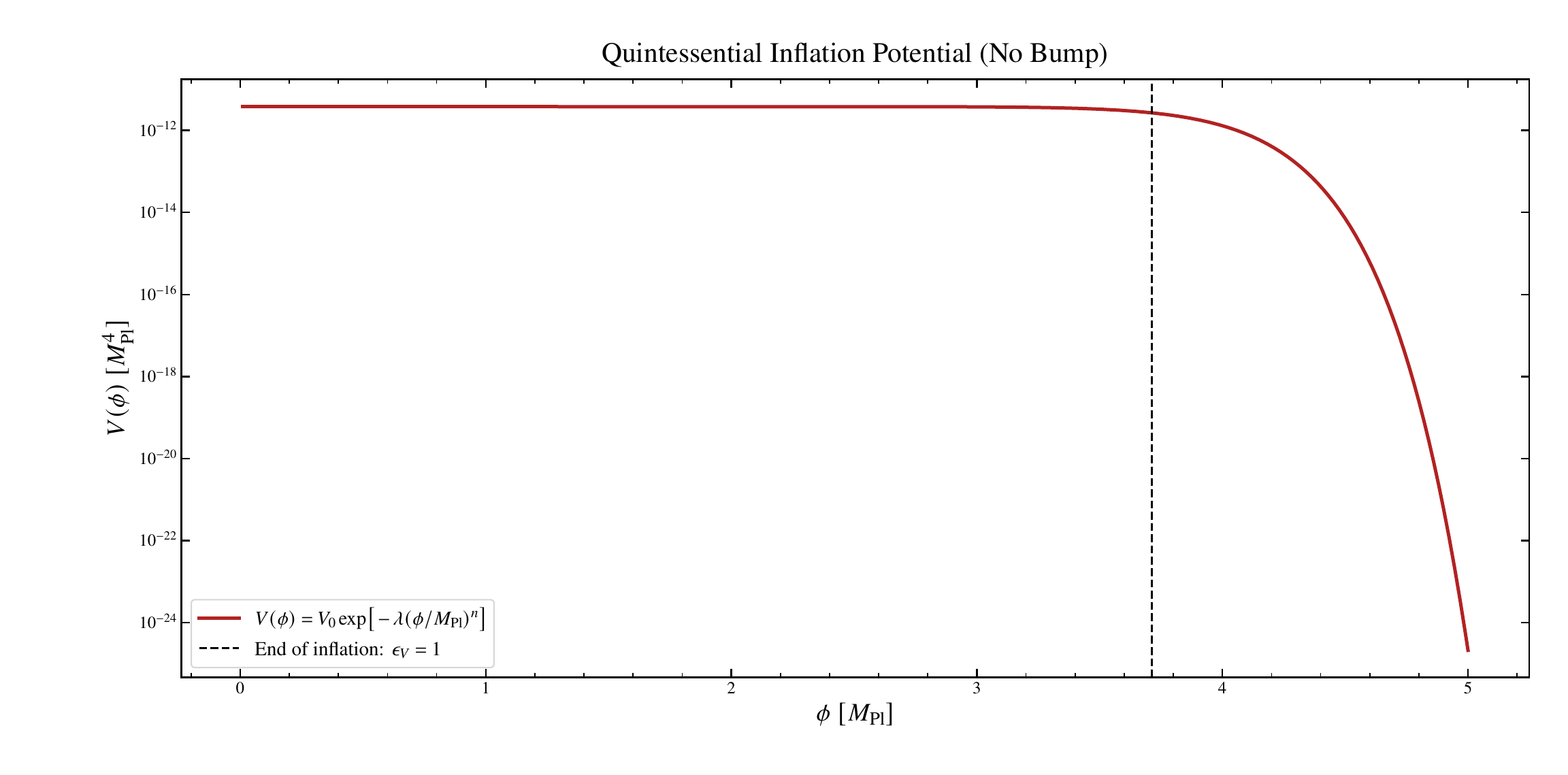}
\caption{Runaway inflationary potential $V(\phi)$ for Eq.~\eqref{eq:V_runaway_refined}. The slow-roll plateau gives way to a steep descent that induces a kinetic-dominated epoch.}
\label{fig:Vphi_plot}
\end{figure*}

Motivated by these developments, it is timely to reassess quintessential inflation in light of the updated ACT DR6 and DESI DR2 \cite{DESI:2024mwx,DESI:2024uvr, ACT:2025fju,ACT:2025tim} measurements, together with current PTA constraints. In the present work, we analyse both smooth and feature-enhanced quintessential potentials using full numerical evolution of the background and cosmological perturbations, avoiding the limitations of analytic slow-roll expressions. We derive updated predictions for the key inflationary statistics $(n_s, r, \alpha_s)$ and examine the associated stochastic gravitational-wave backgrounds over the full frequency interval spanning PTA to planned space-based detectors. Our results highlight the degree to which quintessential inflation remains viable under modern precision cosmology and identify specific parameter ranges in which cosmological gravitational-wave observations serve as decisive tests of the scenario.

The remainder of this article is organized as follows. In Sec.~\ref{sec:quint_no_bump} we present the theoretical setup of our quintessential inflation models and describe our numerical methodology for solving background and perturbation dynamics. Sec.~\ref{sec:constraints} provides constraints derived from the latest ACT DR6 and DESI DR2 datasets and at last we summarize results and outlook in Sec.~\ref{sec:conclusions}.

\section{Quintessential Inflation: A short review}
\label{sec:quint_no_bump}

Quintessential inflation provides a unified cosmological framework in which a single, canonical scalar field accounts for two distinct phases of accelerated expansion in cosmic history: the rapid inflationary expansion in the very early Universe and the presently observed dark energy domination at late times~\cite{Spokoiny:1993kt,Peebles:1998qn,Sahni:2001qp,Sami:2004ic}. Unlike conventional single–field inflationary scenarios that rely on a potential with a minimum triggering inflaton oscillations and standard reheating, quintessential inflation employs a runaway-type potential devoid of a stable minimum. Consequently, the scalar field remains dynamical after inflation ends, eventually serving as the dark energy component responsible for accelerating cosmic expansion today.

The background dynamics of the inflaton $\phi$ in a spatially-flat FLRW Universe are governed by the system
\begin{align}
H^2 &= \frac{1}{3M_{\rm Pl}^2}\left(\frac{\dot{\phi}^2}{2}+V(\phi)\right), \label{eq:QF1}\\
\ddot{\phi} + 3H\dot{\phi} + V_{,\phi} &= 0,
\label{eq:QF2}
\end{align}
where $H$ is the Hubble parameter and overdots denote derivatives with respect to cosmic time. Inflation proceeds while the slow-roll conditions
\begin{equation}
\epsilon_1 \equiv -\frac{\dot{H}}{H^2} \ll 1,
\qquad
\epsilon_2 \equiv \frac{d\ln\epsilon_1}{dN} \ll 1,
\label{eq:slowrollQ}
\end{equation}
hold for a sufficiently long period (typically $50$--$65$ $e$-folds) to resolve the standard cosmological puzzles~\cite{Baumann:2009ds}. Inflation terminates at $\epsilon_1 = 1$, where kinetic energy becomes comparable to potential energy.

A simple and well-studied potential capable of realizing quintessential inflation is the exponential class~\cite{Geng:2015fla,Dimopoulos:2017zvq},
\begin{equation}
V(\phi) =
V_0 \exp\!\left[-\lambda\left(\frac{\phi}{M_{\rm Pl}}\right)^n\right],
\qquad n>1,
\label{eq:V_quint_nonbump}
\end{equation}
where the parameters $(\lambda,n)$ determine the slope and steepness of the potential. For $\phi/M_{\rm Pl}\lesssim\mathcal{O}(1)$ the potential exhibits a sufficiently flat plateau supporting slow-roll, while for larger field values $V(\phi)$ sharply decreases, producing a graceful exit. Normalisation to the amplitude of CMB scalar perturbations fixes $V_0$.

During slow-roll, the curvature perturbation power spectrum is approximated as
\begin{equation}
\mathcal{P}_S(k)
\simeq \frac{H^2}{8\pi^2 M_{\rm Pl}^2 \epsilon_1}\Bigg|_{k=aH},
\label{eq:PSQ}
\end{equation}
leading to observable quantities
\begin{equation}
n_s - 1 \simeq -4\epsilon_1 + 2\epsilon_2,
\qquad
r \simeq 16\epsilon_1.
\label{eq:ns_r_no_bump}
\end{equation}
For suitable choices of $(\lambda,n)$ the model predicts values of $(n_s,r)$ consistent with current CMB constraints from Planck, ACT and DESI~\cite{Planck:2018jri,ACT:2025fju,ACT:2025tim}.\footnote{In this smooth potential realization, $\mathcal{P}_S(k)$ remains nearly scale invariant across all observable scales, precluding substantial small-scale enhancement or primordial black hole formation.}

A distinctive feature of quintessential inflation emerges immediately after inflation ends. The steep decline of the potential rapidly drives the inflaton into a kinetic-dominated (KD) or stiff regime,
\begin{equation}
\rho_\phi \simeq \frac{\dot{\phi}^2}{2} \propto a^{-6},
\qquad
w_\phi \simeq +1,
\end{equation}
which corresponds to the maximum equation-of-state parameter permitted by a canonical scalar field.\footnote{This $w\simeq1$ regime is not naturally realized in standard reheating, where $w$ oscillates about zero. Its existence therefore offers an observational discriminator between quintessential and conventional inflationary models.}  
The scalar energy density redshifts away far faster than radiation or matter, preventing it from over-closing the Universe in the post-inflationary epoch.

Since $V(\phi)$ lacks a minimum, the inflaton does not oscillate and standard reheating is impossible. Instead, radiation must be produced via non-oscillatory mechanisms such as instant preheating~\cite{Bassett:2005xm},
\begin{equation}
\mathcal{L}_{\rm int}
=
-\frac{1}{2}g^2(\phi-\phi_c)^2\chi^2
-h\bar{\psi}\psi\,\chi,
\end{equation}
allowing for rapid transfer of energy when the inflaton crosses $\phi_c$. The resulting reheating temperature must satisfy
\begin{equation}
T_r \gtrsim \mathcal{O}({\rm MeV}),
\end{equation}
to preserve the successful predictions of Big Bang Nucleosynthesis (BBN)~\cite{Figueroa:2018twl}.\footnote{Too early reheating shortens the stiff epoch, erasing the characteristic blue-tilted gravitational-wave signal.}

This non-standard post-inflationary dynamics leaves an imprint in the primordial tensor background. Tensor modes re-entering the horizon during kination are enhanced relative to those entering during radiation domination, producing a blue-tilted spectrum at high frequencies~\cite{Giovannini:1999qj,Riazuelo:2000fc,Ahmad:2019jbm},
\begin{equation}
\Omega_{\rm GW}(k) \propto
\left(\frac{k}{k_r}\right),
\qquad k>k_r,
\label{eq:blue_tilt}
\end{equation}
where $k_r$ marks the horizon scale at the onset of radiation domination. The gravitational-wave amplitude must remain below the BBN bound
\begin{equation}
\Omega_{\rm GW} h^2 \,<\, 1.12 \times 10^{-6},
\end{equation}
thereby constraining the duration of the stiff epoch and the associated reheating dynamics. Future space-based interferometers probing $\mathrm{mHz}$--$\mathrm{kHz}$ frequencies could therefore discriminate quintessential inflation from standard reheating models through this spectral blue tilt.

As the Universe evolves, the inflaton kinetic energy continues to redshift until the shallow tail of the potential once again dominates the dynamics. The field eventually joins the late-time cosmic expansion as dark energy,
\begin{equation}
\rho_\phi \simeq V(\phi),
\qquad w_\phi \rightarrow -1,
\end{equation}
where the potential admits tracker-like solutions that mitigate sensitivity to initial conditions~\cite{Copeland:1997et,Steinhardt:1999nw}.\footnote{Tracker behavior significantly improves phenomenological robustness relative to fine-tuned $\Lambda$CDM-like potentials.}

Thus, quintessential inflation without additional potential features constitutes a minimal and predictive realization of cosmic history: a single degree of freedom drives both primordial inflation and present-day acceleration, introduces a kinetically-driven epoch with a distinctive gravitational-wave signature, and achieves reheating without oscillations while remaining consistent with nucleosynthesis constraints. In later sections, we will investigate extensions of this setup in which localized potential features produce enhanced small-scale power and induced gravitational waves relevant for current PTA measurements.

\section{Numerical Analysis}
\label{sec:constraints}

A complete evaluation of the inflationary performance of the featureless quintessential model requires simultaneous control over the background evolution, the field dynamics across the slow-roll regime, and the resulting inflationary observables. 

The potential responsible for both early-universe inflation and late-time cosmic acceleration takes the form

\begin{equation}
V(\phi)=V_{0}\exp\!\left[-\lambda\left(\frac{\phi}{M_{\rm Pl}}\right)^{n}\right],
\label{eq:V_runaway_refined}
\end{equation}

with benchmark parameters $(V_{0},\lambda,n)=(3.8025\times10^{-12},10^{-9},15)$ selected to ensure correct CMB normalization of the curvature fluctuation amplitude. The exponent $n$ controls the sharpness of the departure from slow roll, with the large value chosen here ensuring a quasi-plateau that sharply transitions into a runaway region. This steep fall-off establishes a canonical kinetic-dominated post-inflationary phase, preventing any inflaton oscillatory reheating and preserving the quintessential nature of the scalar field until present times.

Figure~\ref{fig:Vphi_plot} reveals several critical dynamical properties. The visible plateau at $\phi\gtrsim2.5$ permits a prolonged epoch of accelerated expansion, but the exponential suppression in the region $\phi\simeq2.3$ signals an impending loss of slow-roll behavior. Physically, once the field enters this steep region, its kinetic energy overwhelms the potential energy and the Universe rapidly transitions into a stiff fluid with $w\simeq+1$. The logarithmic scaling in Fig.~\ref{fig:Vphi_plot} emphasizes the many-order-of-magnitude drop in potential energy between horizon-exit scales and the late-time quintessence tail. This provides a direct visual explanation for how a single field can begin by driving primordial inflation and later act as dark energy without introducing ad hoc potentials or separate sectors.

The onset and termination of the inflationary regime can be understood directly from the evolution of the potential slow-roll parameters. These are derived as

\begin{align}
\epsilon_V &= \frac{1}{2}\left(\frac{V_{,\phi}}{V}\right)^{2}, \qquad
\eta_V = \frac{V_{,\phi\phi}}{V}, \\
\xi_V^{2} &= 
\frac{V_{,\phi}V_{,\phi\phi\phi}}{V^{2}},
\qquad
\sigma_V^{3} = 
\frac{V_{,\phi\phi}V_{,\phi\phi\phi\phi}}{V^{2}}.
\end{align}

\begin{figure*}
\centering
\includegraphics[width=\linewidth]{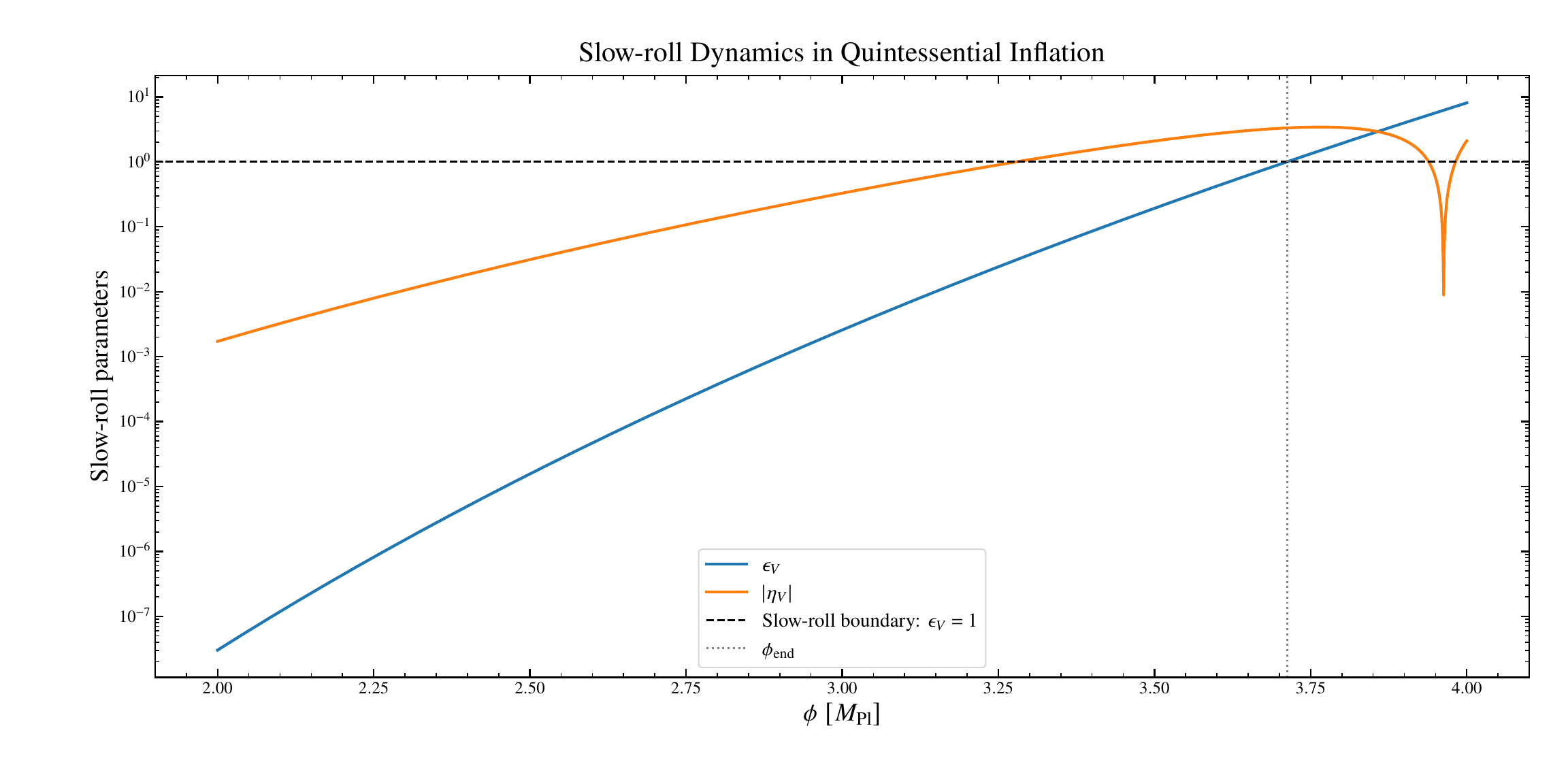}
\caption{Potential slow-roll parameters $\epsilon_V$ and $|\eta_V|$ versus inflaton field value. The end of inflation occurs when $\epsilon_V=1$, driven by curvature variations in the potential.}
\label{fig:slowroll_plot}
\end{figure*}

Figure~\ref{fig:slowroll_plot} confirms that both $\epsilon_V$ and $|\eta_V|$ remain well below unity during most of inflation, ensuring adiabatic suppression of the kinetic contribution. However, as $\phi$ decreases toward $2.39$, slow roll abruptly fails when $\epsilon_V=1$. The sharp rise in $|\eta_V|$ prior to this point reflects an underlying geometric change in the curvature of $V(\phi)$, demonstrating that the end of inflation is driven not only by the growth of the slope but by a rapid variation in the potential curvature. This establishes a predictive relationship where the spectral tilt is largely controlled by $\eta_V$, rather than by $\epsilon_V$.

A mapping between the field evolution and observable cosmological scales is obtained through

\begin{equation}
N(\phi)=\int_{\phi}^{\phi_{\rm end}}\frac{V}{V_{,\phi}}\,d\phi.
\label{eq:Nphi}
\end{equation}

\begin{figure*}
\centering
\includegraphics[width=\linewidth]{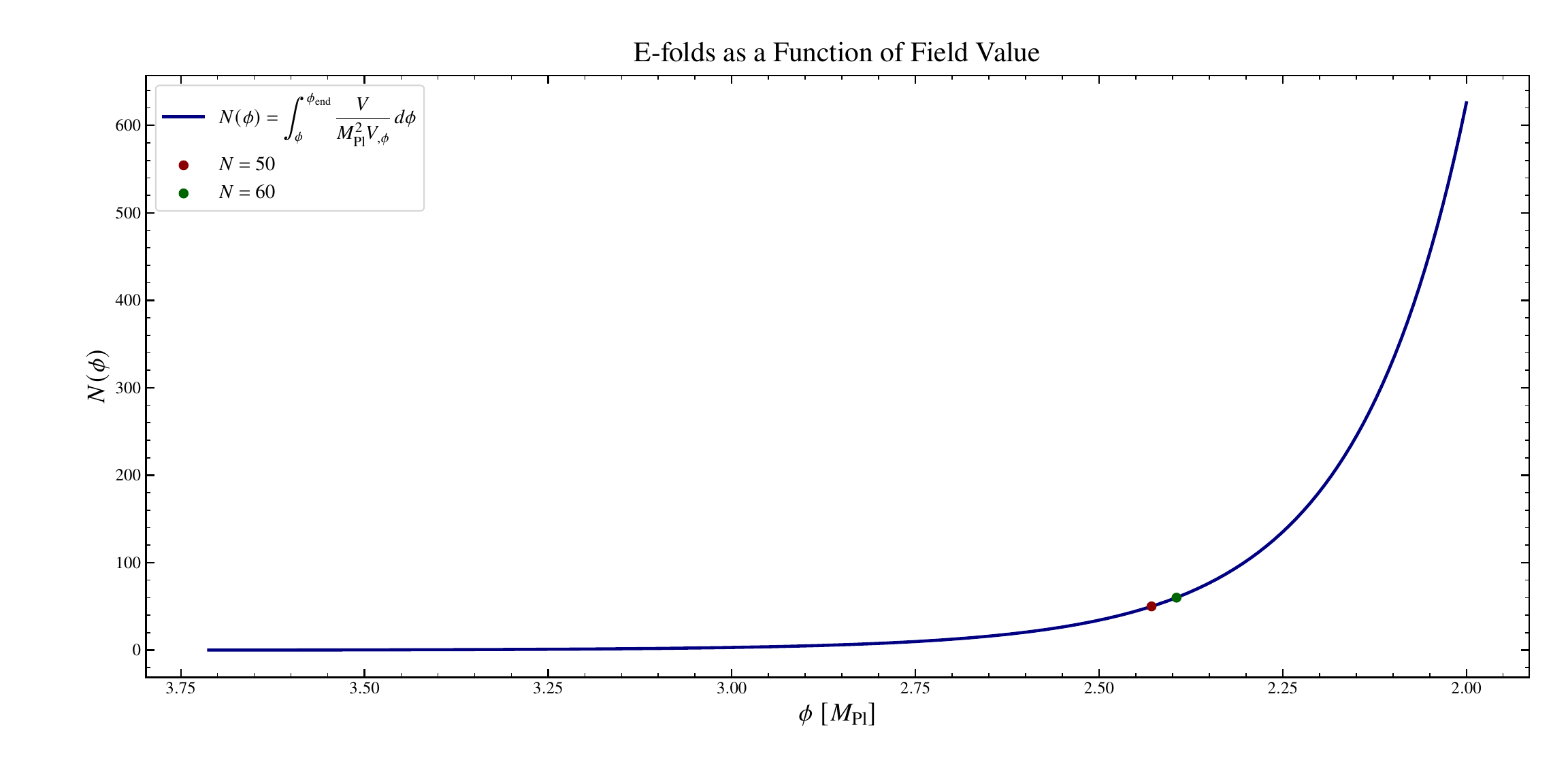}
\caption{E-fold dependence on the inflaton field. The CMB pivot scales probe a limited portion of the potential, enhancing model predictivity and connecting early-time observables to late-time cosmic acceleration.}
\label{fig:Nphi_plot}
\end{figure*}

The resulting function plotted in Fig.~\ref{fig:Nphi_plot} shows that the observable window of inflation is strikingly narrow: modes associated with $N=50$ and $N=60$ exit the horizon at field values only $\Delta\phi\simeq0.03$ apart. This has far-reaching consequences. First, since the relevant physics is confined to such a limited region of the potential, the model acquires powerful predictivity. Second, any measurement that refines the determination of $n_s$ or $\alpha_s$ constrains not only the inflationary plateau but also the post-inflationary evolution, since the field remains dynamically active afterward. Third, because the field displacement during observable inflation is sub-Planckian, quantum gravity corrections remain naturally suppressed, bolstering theoretical robustness.

The inflationary observable predictions can now be evaluated using the standard slow-roll relations

\begin{align}
n_s &= 1 - 6\epsilon_V + 2\eta_V, \qquad
r = 16\epsilon_V, \\
\alpha_s &= 16\epsilon_V \eta_V - 24\epsilon_V^2 - 2\xi_V^2, \\
\beta_s &= -192\epsilon_V^3 + 192\epsilon_V^2\eta_V
          - 32\epsilon_V\eta_V^2 - 24\epsilon_V \xi_V^2 \\ \notag
        &\quad + 2\eta_V\xi_V^2 + 2\sigma_V^3.
\end{align}

The results are provided in Table~\ref{tab:observables}, demonstrating that the spectral tilt remains fully compatible with current CMB constraints. The extremely small tensor-to-scalar ratio places the prediction well below the sensitivity limits of current and near-future CMB B-mode surveys.

\begin{table*}[t!]
\centering
\begin{tabular}{c|c|c}
\hline\hline
Observable & $N=50$ & $N=60$ \\
\hline
$\phi_*/M_{\rm Pl}$ & 2.4286 & 2.3946 \\
$n_s$ & 0.957043 & 0.964241 \\
$r$ & $1.11\times10^{-4}$ & $7.48\times10^{-5}$ \\
$\alpha_s$ & $-8.57\times10^{-4}$ & $-5.94\times10^{-4}$ \\
$\beta_s$ & $2.43\times10^{-2}$ & $1.73\times10^{-2}$ \\
$\epsilon_V$ & $6.94\times10^{-6}$ & $4.67\times10^{-6}$ \\
$\eta_V$ & $-2.15\times10^{-2}$ & $-1.79\times10^{-2}$ \\
\hline\hline
\end{tabular}
\caption{Inflationary predictions for the smooth runaway quintessential model at $N=50$ and $N=60$.}
\label{tab:observables}
\end{table*}

\begin{figure*}
\centering
\includegraphics[width=\textwidth]{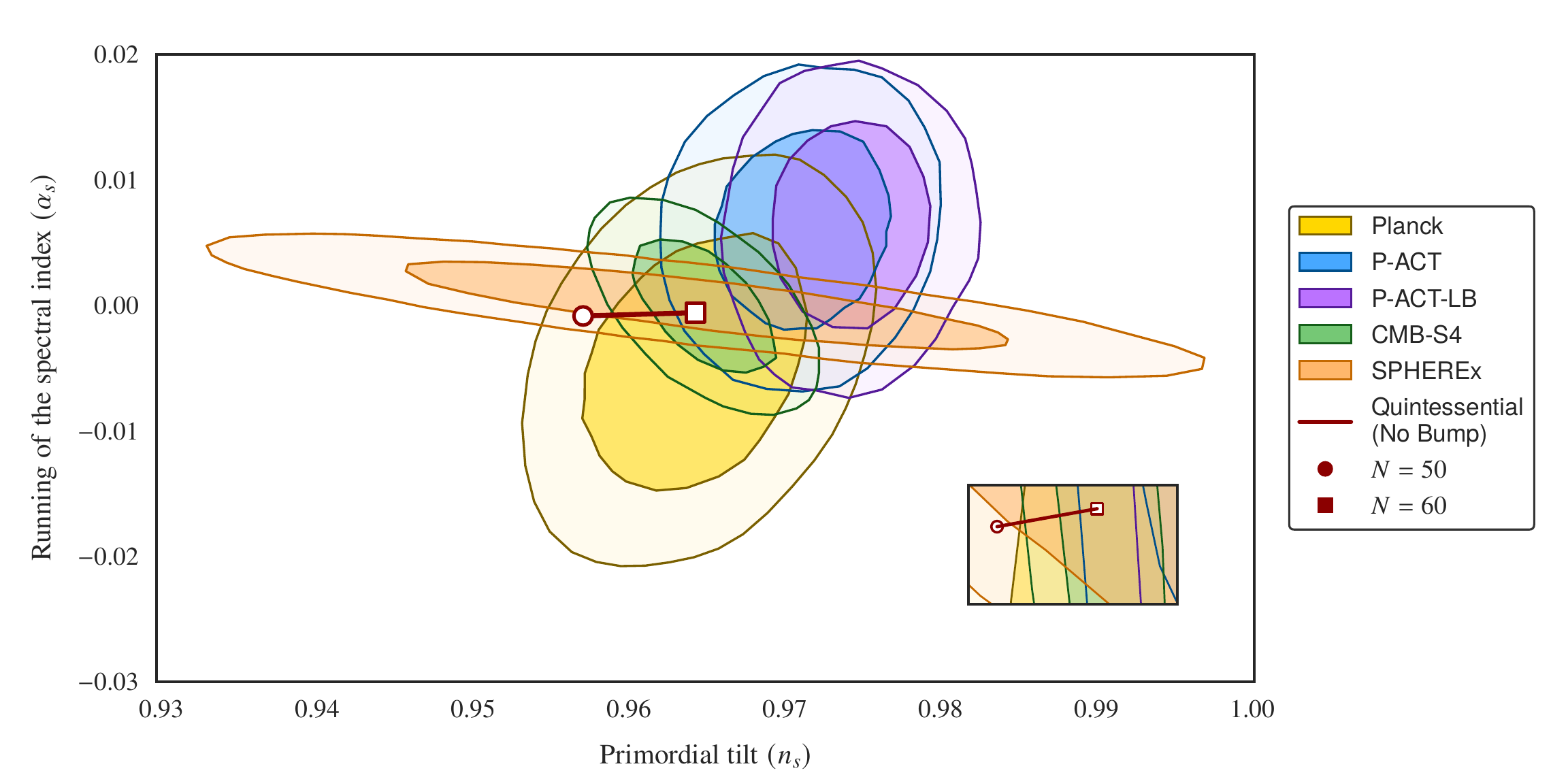}
\caption{Model predictions in the $(n_s,\alpha_s)$ plane compared with present and forecast CMB constraints. The theoretical trajectory traces a narrow curve consistent with the observed degeneracy direction.}
\label{fig:ns_alpha_quint}
\end{figure*}

The implications of the above values can be appreciated by placing them in the $(n_s,\alpha_s)$ parameter space as shown in Fig.~\ref{fig:ns_alpha_quint}. The trajectory described by the model aligns with the dominant degeneracy direction of current CMB constraints. This alignment indicates that improving measurements of $n_s$ and $\alpha_s$ will tighten the allowed inflationary window orthogonal to this trajectory, providing a powerful test of the scenario. The predicted negative running emerges directly from the exponential steepening of the potential and therefore provides a probe of the physics responsible for the post-inflationary stiff epoch, establishing a direct link between early-universe observables and the late-time acceleration mechanism.

\begin{figure*}
\centering
\includegraphics[width=\linewidth]{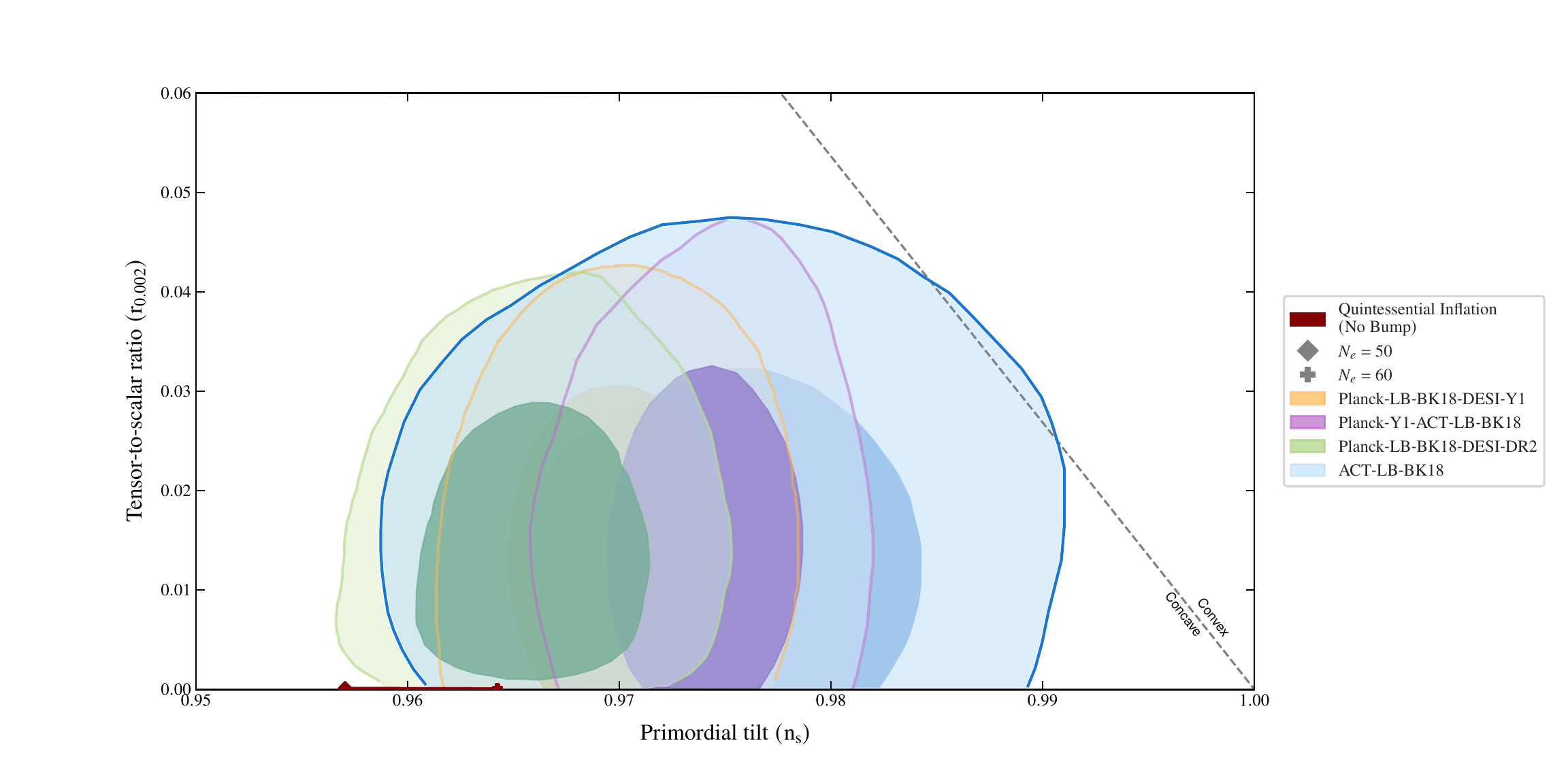}
\caption{Predictions in the $(n_s,r)$ plane are fully consistent with current bounds. Despite the suppressed tensor amplitude at CMB scales, high-frequency gravitational waves provide an essential observational window into the post-inflationary stiff epoch.}
\label{fig:ns-r}
\end{figure*}

A final perspective comes from tensor modes, shown in Fig.~\ref{fig:ns-r}. The tiny values of $r$ obtained here confirm that CMB-based tensor searches will not discover gravitational waves from inflation in this model. Yet this low amplitude is not a conclusion of observational invisibility but a shift in experimental target. The post-inflationary stiff phase greatly enhances high-frequency gravitational-wave power, generating a blue-tilted spectrum that lies within reach of future interferometric detection. Crucially, this connects gravitational-wave probes to the reheating history and the viability of quintessential models as dark energy candidates, allowing tensor phenomenology to extend well beyond its traditional CMB role.

Taken together, these results demonstrate that the smooth runaway quintessential potential naturally produces realistic inflationary predictions, remains compatible with all current CMB datasets, and makes testable predictions regarding future refinements in primordial spectral measurements. Most importantly, it establishes that the physics driving cosmic acceleration today is encoded in the same scalar field responsible for inflation, enabling a single degree of freedom to unify both acceleration epochs while remaining falsifiable through its gravitational-wave signature across a broad frequency range.

\section{Results and Discussion}
\label{sec:conclusions}

The analysis presented in Sec.~\ref{sec:constraints} demonstrates that the featureless quintessential inflation model under consideration yields a highly predictive inflationary sector in excellent agreement with present CMB observations. The computed values of $(n_s, \alpha_s, r)$ for the benchmark choices $N=50$ and $N=60$ lie entirely within the currently allowed regions from Planck, ACT, and related datasets, as illustrated in Figs.~\ref{fig:ns_alpha_quint} and \ref{fig:ns-r}. The inflationary predictions exhibit a smooth dependence on the number of $e$-folds, reinforcing the consistency of the slow-roll regime supported by the runaway exponential potential.

The scalar spectral index is found to be mildly red-tilted, with $n_s \lesssim 0.965$ for $N=60$, while the tensor-to-scalar ratio remains strongly suppressed at the level $r \sim \mathcal{O}(10^{-4})$. These values place the model deep within the concave inflationary region preferred by current data. Importantly, the suppression of primordial tensor modes is not a generic signature of low-energy inflation but emerges here through the extreme flatness of the potential around the CMB pivot scales. Such behavior distinguishes quintessential inflation from plateau models that predict larger values of $r$ nearer the observable threshold.

Despite its low tensor amplitude at horizon scales, the model remains far from observational irrelevance in the gravitational-wave sector. The post-inflationary stiff phase, characterized by $\rho_\phi \propto a^{-6}$ and $w_\phi \simeq +1$, amplifies high-frequency tensor modes relative to the nearly scale-invariant background produced during inflation. This effect, intrinsically linked to the quintessential nature of the scalar field, gives rise to a blue-tilted spectrum of primordial gravitational waves that may lie within reach of future interferometric missions operating in the millihertz to kilohertz frequency range. Consequently, gravitational-wave astronomy offers a non-CMB pathway for probing early-universe dynamics in quintessential inflation, particularly in scenarios where reheating is mediated by non-oscillatory processes.

A second noteworthy aspect arises from the compactness of the inflationary field domain explored by observable modes. The small variation $\Delta\phi \approx 0.03 M_{\rm Pl}$ between $N=50$ and $N=60$ shows that the phenomenology of the CMB does not depend on extended excursions across field space. This enhances the robustness of the model with respect to potential quantum-gravity corrections and strongly constrains permissible deviations in the inflaton potential. As experimental precision improves, particularly through next-generation surveys such as CMB-S4 \cite{Abazajian:2016yjj}, LiteBIRD \cite{LiteBIRD:2022cnt}, and SPHEREx \cite{Dore:2014cca}, the narrow theoretical trajectory traced in the $(n_s,\alpha_s)$ and $(n_s,r)$ planes will be subject to increasingly discriminating tests.

Measurements of $\alpha_s$ are especially promising in this context. The predicted negative running is modest yet appreciable, directly induced by the exponential steepening that governs the transition to quintessence. Such behavior is absent in many plateau and attractor models that feature negligible running. Hence, any significant detection or sharp upper bound on $\alpha_s$ in the coming decade will convey direct information on both the inflationary exit dynamics and the onset of kination.

In summary, the smooth quintessential inflation scenario considered here achieves three key outcomes that collectively elevate its scientific relevance. First, it provides a unified description of early- and late-time cosmic acceleration without the need for multiple scalar sectors. Second, it produces inflationary observables entirely consistent with current constraints while generating future predictions along a single, well-defined path in parameter space. Third, and uniquely, it preserves observational access to the post-inflationary era through a blue-tilted stochastic gravitational-wave background extending into the interferometric frequency range.

These features ensure that quintessential inflation remains a testable and falsifiable framework. While current CMB probes strongly support its inflationary predictions, upcoming gravitational-wave surveys will play a central role in verifying the post-inflationary kinetic-dominated epoch that distinguishes this class of cosmological models from conventional reheating scenarios. The full picture emerging from these complementary investigative channels strongly motivates further theoretical and observational scrutiny of quintessential inflation as a compelling candidate for a unified cosmological history.

\section*{Acknowledgement}
SC would like to thank M. Sami for the useful discussions.

\bibliography{RefsGWB}
\bibliographystyle{utphys}

\end{document}